\documentclass[%
reprint,
showpacs,
 amsmath,amssymb,
 aps,
prl,
]{revtex4-1}

\usepackage{graphicx}
\usepackage{dcolumn}
\usepackage{bm}
\newcommand{\be}{\begin{equation}}
\newcommand{\ee}{\end{equation}}
\newcommand{\bea}{\begin{eqnarray}}
\newcommand{\nn}{\nonumber}
\newcommand{\eea}{\end{eqnarray}}

\begin{document}

\preprint{APS/123-QED}

\title{Effectively universal behavior of rotating neutron stars in general relativity makes them even simpler than their Newtonian counterparts}

\author{George Pappas}
\email{gpappas@sissa.it}
\affiliation{SISSA, Via Bonomea 265, 34136 Trieste, Italy}
\author{Theocharis A. Apostolatos}%
\email{thapostol@phys.uoa.gr}
\affiliation{Section of Astrophysics, Astronomy, and Mechanics,
Department of Physics, University of Athens, Panepistimiopolis Zografos GR15783,
Athens, Greece}%

\date{\today}

\begin{abstract}

Recently it was shown that slowly rotating neutron stars exhibit an interesting correlation between their moment of inertia $I$, their quadrupole moment $Q$, and their tidal deformation Love number $\lambda$ (the I-Love-Q relations), independently of the equation of state of the compact object. 
In the present work a similar, more general, universality is shown to hold true for all rotating neutron stars within General Relativity; the first four multipole moments of the neutron star are related in a way independent of the nuclear matter equation of state we assume. By exploiting this relation, we can describe quite accurately the geometry around a neutron star with fewer parameters, even if we don't know precisely the equation of state. Furthermore, this universal behavior displayed by neutron stars, could promote them to a more promising class of candidates (next to black holes) for testing theories of gravity.  

 \end{abstract}



\maketitle



{\it Introduction.---}
%
Neutron stars (NSs) constitute a class of the most interesting laboratories for studying extreme physics in nature. The physical processes taking place in these astrophysical objects involve on the one hand strong gravity effects, second only to the inexorable gravitational effects of astrophysical black holes, and on the other hand microscopic effects that are related to the properties of matter in densities that exceed nuclear density. Consequently, NSs offer a valuable tool for broadening our understanding of gravity, by testing the predictions of the established theory of General Relativity (GR) or alternative theories of gravity, as well as our understanding of the microphysics ruling matter at densities as high as those found at the centre of NSs. This effort can be hindered by the complexities of the microphysics that enters the description of NSs and the way these complexities propagate to the gravitational aspects of the problem. Thus understanding NSs structure and gravitational field is of paramount importance.

Recently, it was shown in \cite{Yagi-Yunes} that for slowly rotating NSs there exist some interesting and unexpected universal relations between the normalized moment of inertia $\bar{I}\equiv I/M^3$, the normalized quadrupole $\bar{Q}\equiv Q/(M^3j^2)$ (where $M$ is the mass, $j\equiv J/M^2$ is the spin parameter, and $J$ is the angular momentum), and the normalized Love number $\bar{\lambda}\equiv \lambda/M^5$, that is, relations that are independent of the specific equation of state (EoS) assumed to describe the NS matter. It was argued that one could use these universal relations in the gravitational wave analysis of NS/BH or NS/NS binary inspirals to break the degeneracy, showing up at 2PN order, between the quadrupole moment and the spin-spin interaction terms of the two bodies, and thus measure the properties of each member individually. Moreover, it was argued that these relations could be used to perform tests of the theory of gravity (i.e., to distinguish between NSs in different theories of gravity and NSs within GR), since any EoS uncertainties and their subsequent gravitational effects could be eliminated. 

However, it was shown in \cite{Donevaetal}, that these relations do not hold for rapidly rotating NSs, where the rotation was parameterized with the rotation frequency of the NS. Even for moderate rotation rates, i.e., rotation frequencies above a few hundred Hz, deviations from the universal relation of \cite{Yagi-Yunes} between the normalized quadrupole and the normalized moment of inertia ($\bar{I}-\bar{Q}$ relation) start showing up. Therefore, a frequency dependence is introduced beyond the slow-rotation approximation. This dependence is weaker as one approaches the maximum mass limit for NSs, since then NSs behave more like black holes. The loss of universality at large rotation rates though, does not lessen its significance, since NSs involved in inspiraling binaries are not expected to rotate really fast\cite{Ferrari}. The effects of magnetic fields have also been explored in \cite{Luciano}.

While studying the relativistic multipole moments of compact objects \cite{PappasMoments} in order to use them as parameters to construct analytic spacetimes that are mimicking the geometry around such objects \cite{twosoliton}, we found an unexpected universal connection between the first four relativistic multipole moments, that was independent of which realistic EoS was used to construct the NS models. Namely, the values of the reduced parameters, $j$, $\sqrt{-q}\equiv \sqrt{-Q/M^3}$, and $\sqrt[3]{-s_3}\equiv\sqrt[3]{- S_3/M^4}$, where $Q$ is the quadrupole and $S_3$ is the spin octupole of the Hansen-Geroch moments \cite{Gero,Hans}, were found to lie on a single, almost planar, surface independent of the EoS. This is analogous to the reduced moments of Kerr black holes which lie on a line. Effectively this universality could be considered as a ``no hair'' property of neutron stars, resembling that of black holes.

The significance of this result has two aspects. On the one hand, the observed behavior implies that the extra degrees of freedom that are encoded in realistic EoSs (realistic EoSs can be described as piecewise polytropes with several parameters varying from the one EoS to the other) are irrelevant to the relations between the multipole moments. Thus one could obtain a simple description of the spacetime around NSs and its stationary properties, independently of the details of the EoS. Furthermore, the relations between the moments could be used to extract information from astrophysical systems in the gravitational-wave window, as it is elaborated in \cite{Yagi-Yunes}, as well as in the electromagnetic window, where the moments are the parameters to be extracted from the observations (see for example \cite{PappasQPOs,psaltis2,psaltis3,psaltis4}). 

On the other hand, this behavior of NSs in GR offers, at least in principle, the possibility of new tests of gravity. One possible avenue of exploration could be to attempt an observational verification of this behavior of NSs, although such an endeavor will be really challenging. Another aspect, worth exploring, is whether NSs in alternative theories of gravity display similar behavior. That, if nothing else, would be an interesting theoretical investigation in the properties of alternative theories (some first results have been produced in Chern-Simons \cite{Yagi-Yunes} and in Eddington-inspired Born-Infeld \cite{Sham} gravity). However, it should be emphasized that any comparison between GR and alternative theories of gravity should be preceded by a careful construction of appropriate corresponding quantities in other theories. 

Finally, we should note that when a scale, such as the mass or the rotational frequency of the star, is introduced in the description, the universality of the reduced moments between different EoSs breaks and individual EoSs can be discerned.    

The rest of this letter is organized as follows. First we present the setup of our analysis and briefly discuss some aspects of the Newtonian theory to gain some insight on the behavior of the moments. Then we present our results regarding the universal behavior, followed by a comparison with the previously established I-Love-Q universality. We conclude with a short discussion on how one can break the degeneracy between the different EoSs. 

It is worth mentioning that after the submission of this work, further corroborating results have appeared \cite{Chakrabarti,Stein}.

{\it Numerical models and the parameter space.---}
%
%
For our analysis we have constructed NS models using the \texttt{RNS} numerical code of \cite{Sterg}. For these models we computed the multipole moments according to the prescription presented in \cite{PappasMoments} (some first results for the moments of NSs have been presented in \cite{Pappas2012}). The EoSs that we have used  \cite{AandB,WFF,SLy4,NegeleVautherin,BPS,APR,FPS,balb1} to construct the NS models are presented in Table I of \cite{suppl}. Apart from the nuclear matter EoSs, we have also used the two proposed EoSs in \cite{SLB} as inferred by Bayesian analysis from astrophysical observations of type I X-ray bursters with photospheric radius expansion and from thermal emission from quiescent low-mass X-ray binaries. We will generally refer to all these EoSs as ``the realistic" EoSs. For our analysis we have used models from a little less than $1 M_\odot$, up to the maximum stable mass for each EoS. The masses are expressed in geometric units ($1M_{\odot}=1.477 \textrm{km}$). 

For each of these models (see details in \cite{suppl}) we obtain a multitude of physical parameters, among which are the first non-zero multipole moments, i.e., the mass $M$, the angular momentum $J$, the mass quadrupole $Q$, and the spin octupole $S_3$. As it was shown in \cite{poisson,PappasMoments}, the higher reduced moments scale with some appropriate power of the spin parameter $j$, i.e.,  
\bea 
q &=& -a(M, \textrm{EoS})j^2,\label{moments1}\\
s_3 &=& -\beta (M, \textrm{EoS})j^3,\label{moments2}
\eea
where the coefficients $a$ and $\beta$ depend in general on the mass of the NS and the EoS. Here, we will consider the three dimensional parameter space of $j$, $\sqrt{a}\equiv \sqrt{-q}/j$, and $\sqrt[3]{\beta}\equiv \sqrt[3]{-s_3}/j$. 

We should note at this point that, as it was shown in \cite{twosoliton}, by using the Two-Soliton analytic spacetime \cite{Manko}, the first four non-zero moments are capable of producing an accurate description of the spacetime exterior of NSs, while the first moments could in principle be measured by using for example quasi-periodic oscillations as probes of the geometry around NSs \cite{PappasQPOs}, or other methods \cite{psaltis2,psaltis3,psaltis4}. Therefore these four moments constitute in practice all the necessary information for the description of the spacetime around a NS and are in principle accessible by astrophysical observations.

Before presenting our results, it would be useful if we could gain some insight of what to expect. For slowly rotating Newtonian polytropes (including deformations of $O(\Omega^2)$) one can show that the multipole moments scale as,
\bea 
J\propto \rho_c^{\frac{5-3n}{2n}} \Omega_{\star}\Rightarrow j\propto \rho_c^{\frac{-1-n}{2n}} \Omega_{\star}\propto \rho_c^{-\frac{1}{2n}} \left(\frac{\Omega_{\star}}{\Omega_K}\right), \label{newtonJ}\\
Q\propto \rho_c^{\frac{5(1-n)}{2n}} \Omega_{\star}^2\Rightarrow q\propto \rho_c^{\frac{-2-n}{n}} \Omega_{\star}^2\propto \rho_c^{-\frac{2}{n}} \left(\frac{\Omega_{\star}}{\Omega_K}\right)^2 ,\label{newtonQ}\\
S_3\propto \rho_c^{\frac{7(1-n)}{2n}} \Omega_{\star}^3\Rightarrow s_3\propto \rho_c^{\frac{-5-3n}{2n}} \Omega_{\star}^3\propto \rho_c^{-\frac{5}{2n}} \left(\frac{\Omega_{\star}}{\Omega_K}\right)^3\!,\label{newtonS3}
\eea
where $\rho_c$ and $\Omega_{\star}$ are the central density and the rotation frequency of the star respectively, $n$ is the polytropic index of the polytrope, $P=K\rho^{1+1/n}$, and $\Omega_K$ is the Kepler frequency, which is $\Omega_K\propto\rho_c^{1/2}$ (we should also note that $M\propto\rho_c^{\frac{3-n}{2n}}$ and $R\propto\rho_c^{\frac{1-n}{2n}}$, and that all these relations hold also for GR polytropes, see \cite{Tooper}). From these expressions we can see that the parameters $\sqrt[3]{\beta}$ and $\sqrt{a}$ of equations (\ref{moments1},\ref{moments2}) (these equations are valid for slow rotation, as one can infer from the above formulae) can be related as follows,
\be \left. \begin{array}{c}
  \sqrt{a}=\sqrt{-q}/j\propto\rho_c^{-1/2n}  \\
  \sqrt[3]{\beta}=\sqrt[3]{-s_3}/j\propto\rho_c^{-1/3n} \\ 
\end{array}\right\} \Rightarrow \sqrt[3]{\beta}\propto (\sqrt{a})^{2/3}.\label{unirelation} 
\ee
We will use this Newtonian behavior as a guide to analyze the behavior of NSs in GR. Namely, we will seek a relation between the parameters $\sqrt{a}$, and $\sqrt[3]{\beta}$ in GR of the form $ \sqrt[3]{\beta}=B (\sqrt{a})^{\nu}$; if we had taken the Newtonian result at face value we would have $\nu=2/3$, while $B$ would depend on the EoS.

\begin{figure}
\centering
\includegraphics[width=.4\textwidth]{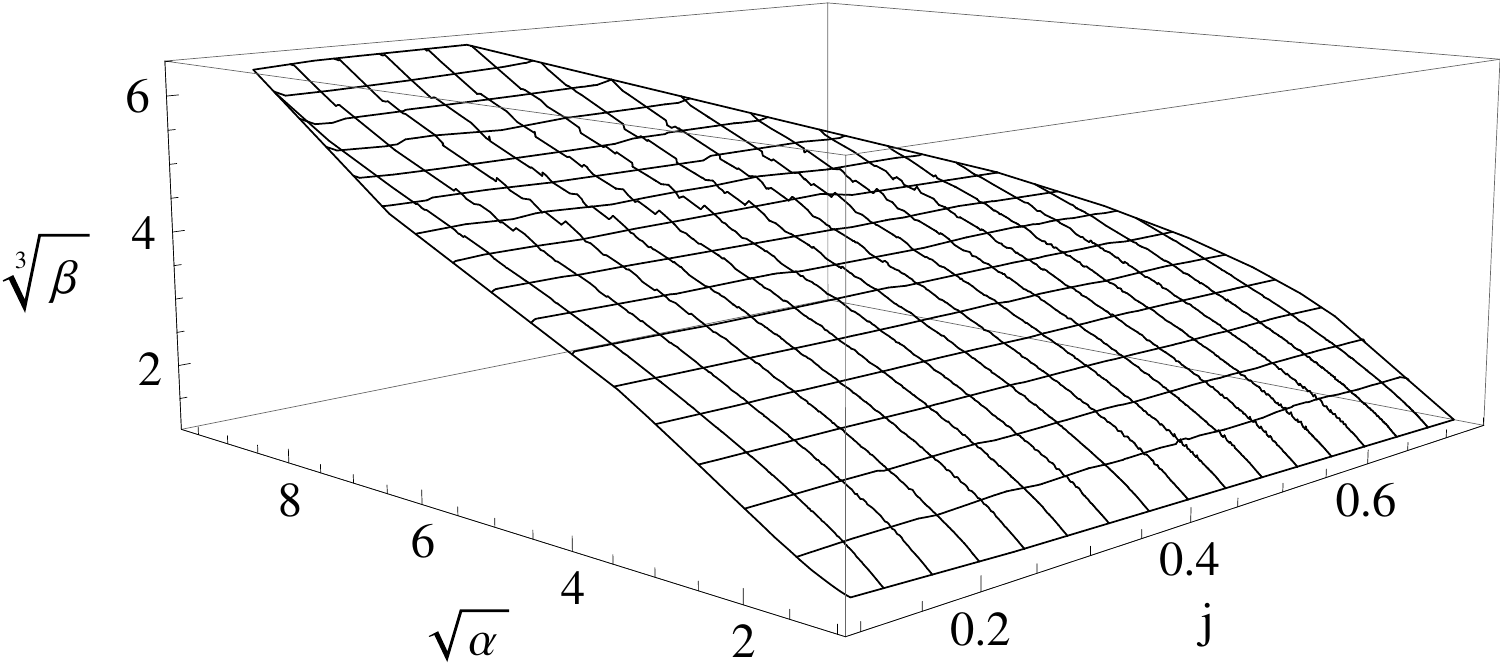}
\vspace{-5pt}
\caption{Plot of the surface formed by all the models constructed using realistic EoSs. All the models fall on one very well defined surface.}
\vspace{-15pt}
\protect\label{UniSurface}
\end{figure}

{\it Universal behavior.---}
%
The question arising now is, what is the behavior of the moments for the realistic EoSs in GR and do different EoSs display different behavior? Surprisingly, the answer is that all realistic EoSs behave as a single polytrope-like EoS with respect to the relations between the multipole moments, forming a very well defined surface in the parameter space of $(j,\sqrt{a},\sqrt[3]{\beta})$. This surface is shown in Figure \ref{UniSurface}. The central density of the NS models increases as $\sqrt{a}$ decreases. The right edge of the surface approaches the line of $\sqrt{a}=\sqrt[3]{\beta}=1$, which corresponds to Kerr black holes. Notice that $\sqrt[3]{\beta}$ appears to have almost no dependence on $j$; thus it can be fitted very well with a function of the form, 
\be \sqrt[3]{\beta}=B (\sqrt{a})^{\nu}, \label{unirelation2} \ee 
with fitting parameters, $B=1.17$, and $\nu=0.74$. A better fit can be obtained if we choose the function, $ \sqrt[3]{\beta}=A+B (\sqrt{a})^{\nu}$, with fitting parameters, $A=-0.36$, $B=1.48$, and $\nu=0.65$, which is surprisingly close to 2/3. An even more accurate fit (better than 2\%) is also given in \cite{suppl}. This accuracy is indicative of how well all NS models fall on a surface.

The fact that all the models seam to fall on the same surface is intriguing and suggests that the moments, and more specifically the parameters $a$ and $\beta$ of equations (\ref{moments1},\ref{moments2}), depend on only two parameters (rotation and central density) without any significant dependence on any other EoS related parameter. Otherwise, the models would not occupy a surface and would be more scattered in 3D.
 
%
%

The explanation of why different EoSs work the same way with respect to the relations between the reduced moments can be implied by the discussion of the moments of Newtonian polytropes. The coefficients that have been omitted in the expressions (\ref{newtonJ},\ref{newtonQ},\ref{newtonS3}), which make them exact equations, depend on the particular EoS; in the Newtonian polytropic case, the coefficients depend on the constant $K$ of the polytropic equation, the polytropic index $n$, and the radial profile of the specific Lane-Emden function. Actually, $K$ does not enter at all in the reduced quantities $j,~q$, and $s_3$, since it has been scaled out by their definition as dimensionless quantities. The remaining characteristics of a specific EoS are anticipated to introduce a dependence of the parameter $B$ in equation (\ref{unirelation2}) on the EoS. This dependence should be carried over from the Newtonian theory to GR. But then, why do different realistic EoSs lead to the same $B$ parameter? One could possibly suspect that there is a conspiracy in the calculation of $B$ so as to end up having the same value regardless of the EoS. Alternatively, the answer could be simply that all realistic EoSs behave as a single quasi-polytropic EoS, with respect to moment determination. Next we will explain why we believe that the latter explanation is the right one.

Most nuclear matter EoSs behave as if they have an effective polytropic index $n$ which is close to 1 (see for example \cite{LattimerPrakash}), a property exhibited by an almost indifference of the radius of the star to its central density (see Figure 1 in \cite{suppl}). In order to test our hypothesis that the observed behavior is due to the fact that all realistic EoSs behave more or less as a single quasi-polytrope, we constructed families of rotating fluid spheres with various polytropic indices and plotted these models in the same parameter space. 

\begin{figure}
\includegraphics[width=.40\textwidth]{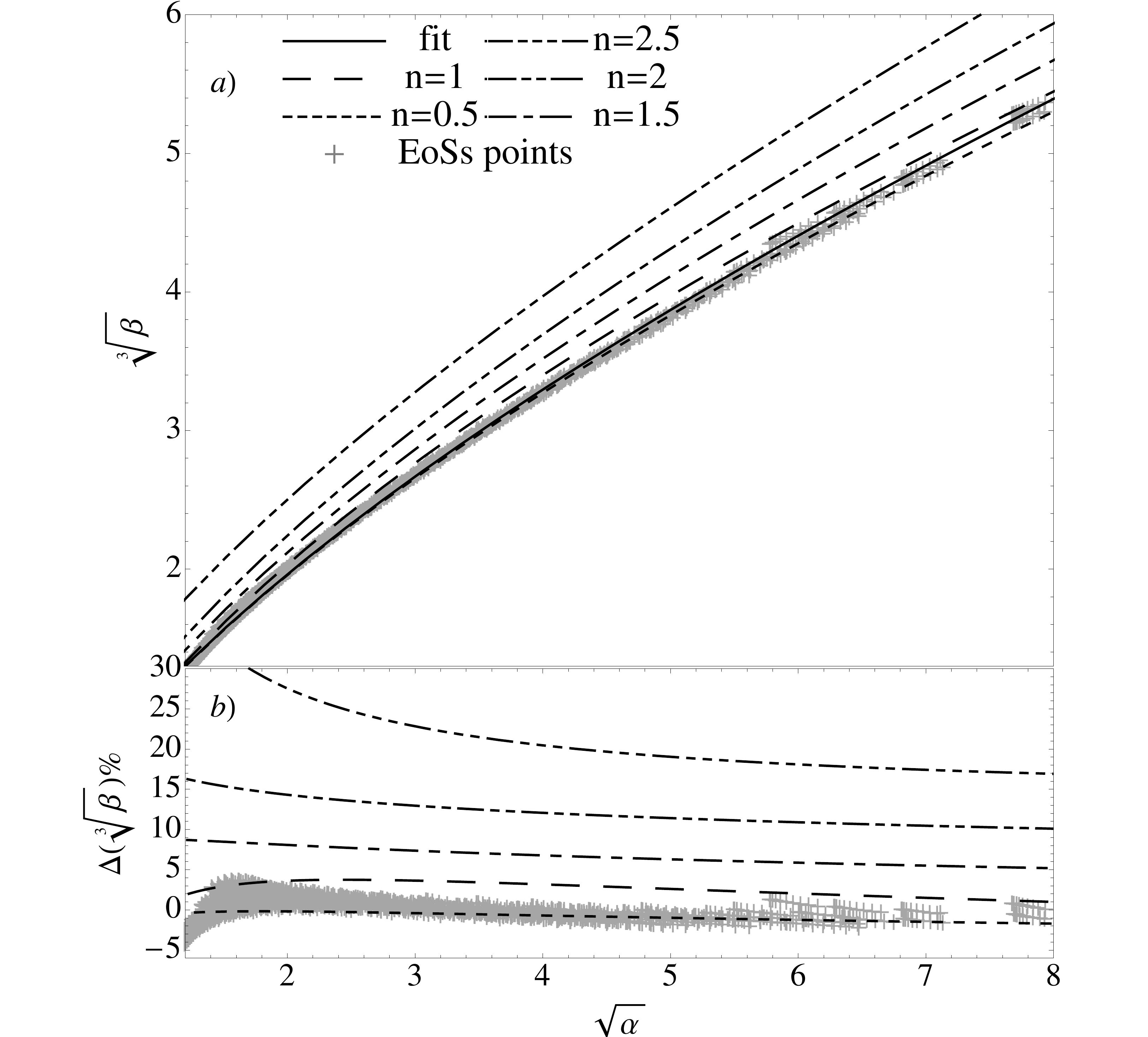}
\vspace{-5pt}
\caption{
Plot a) is the projection on $j=0$ of the best fit surfaces for the realistic EoSs and the polytropic EoSs. Plot b) is the relative difference \% of the polytropic best fit surfaces from the realistic EoSs best fit surface. 
}
\vspace{-15pt}
\protect\label{CompSurface}
\end{figure}

The result is that all the polytropic models form surfaces, similar to that formed by realistic EoSs, which are distinct for every $n$ and are described by a relation between $\sqrt[3]{\beta}$ and $\sqrt{a}$ that is almost independent of $j$ and can be given in the form of equation (\ref{unirelation2}). In Figure \ref{CompSurface}(a), we show the best fit curves for the polytropes, which are essentially the projections of the surfaces they form, on the $j=0$ plane (the parameters for these curves are given in Table II in \cite{suppl}). Figure \ref{CompSurface}(b) shows, the relative difference between the surfaces of the polytropic EoSs and the surface of the realistic models. From this plot one can see that the surfaces of the various polytropes are indeed distinct and are not intersecting with each other.

Of course a particular polytrope is expected to form a surface in the $(j,\sqrt{a},\sqrt[3]{\beta})$ space, since polytropic solutions depend on two parameters, central density and rotation. Also the multipole moments of all polytropes follow relations of the form (\ref{moments1},\ref{moments2}), just as the realistic EoS models do. This is a consequence of the fact that both $\sqrt[3]{\beta}$ and $\sqrt{a}$ are almost independent of $j$. This behavior was not anticipated in GR, since this is not the case at least in Newtonian theory of rapidly rotating stars of uniform density, as it is noted in \cite{poisson}. One could conclude that, from that point of view, the GR rotating fluid spheres essentially behave as slowly rotating Newtonian fluid spheres (including deformations $O(\Omega^2)$). This behavior has also been noted in \cite{Butt}, where it was found that when comparing GR and Newtonian polytropic models with $n<2.5$ (for models of equal angular velocity and central rest mass density), the relativistic models are more spherical. Additionally, since each polytrope falls on a distinct surface (Figure \ref{CompSurface}), the determination of the multipole moments, or to be more specific, of the coefficients $a$ and $\beta$, can distinguish different polytropic EoSs.

The comparison of the realistic EoSs against the polytropes in Figure \ref{CompSurface} suggests that the former ones behave, with respect to their multipole moments, very much like a polytrope with an index $n \simeq 1$. More precisely, there seems to be a monotonic variation of realistic EoSs' behavior between $n\sim0.5$ for high density models, which behave more like fluid spheres of uniform density (these models have multipole moments close to the moments of Kerr black holes and are more compact), and $n\sim1$ for lower density models. If we had included in our analysis very low mass models with very low central densities, then these models would tend towards the $n\sim1.5$ curve, corresponding to NSs with EoSs closer to that of non-relativistic degenerate neutrons, with larger radii and larger multipole moments. Furthermore, the rapidly rotating stars of low mass, being less compact, are expected to be less relativistic and therefore behave less like their slowly rotating Newtonian counterparts and more like rapidly rotating Newtonian stars (we have seen indications of that behavior in some models of very low masses).

To summarize, one could say that the observed universal behavior shows up because on the one hand all the realistic EoSs behave as the same quasi-polytrope, with an effective polytropic index a little lower than 1, and on the other hand the NSs within GR, even when they are rapidly rotating, behave as slowly rotating Newtonian fluid spheres.

{\it Connection to I-Love-Q.---}
%
%
At this point we would also like to explore a possible connection of our results with those in \cite{Yagi-Yunes} and \cite{Donevaetal}. Thus we plotted all our NS models in the parameter space of $(j,\sqrt{a},\sqrt{\bar{I}})$, where $a$ is by definition the $ \bar{Q}$ of \cite{Yagi-Yunes, Donevaetal}) and $\bar{I} \equiv I/M^3$ ($I$ is the moment of inertia of the star). This is similar to the parameter space used in \cite{Donevaetal}, i.e., the parameter space of $(f, \bar{Q}, \bar{I})$, where $f$ is the rotation frequency of the NS, though not exactly equivalent. Again all NSs form a single surface independent of the EoS. A plot of this surface is depicted in Figure 6 in \cite{suppl}. The surface has some non-negligible dependence on the spin parameter $j$, so we have attempted to fit it using a simple polynomial function of the form,
\be
\sqrt{\bar{I}}=A_1+ A_2 (\sqrt{a}-\xi_0) + A_3 (\sqrt{a}-\xi_0)^2, \label{iqfit}
\ee
where, $A_2=B_1+B_2 j +B_3 j^2$, $A_3=C_1+C_2 j +C_3 j^2$, and $\xi_0$ is a constant. 
The best fitting values for the parameters are, $A_1=2.16$, $\xi_0=1.13$, $B_1=0.97$, $B_2=-0.14$, $B_3=1.60$, $C_1=0.09$, $C_2=0.23$, and $C_3=-0.54$. With these parameters the fit is better than 1\%. Our results are equivalent to those in \cite{Donevaetal}, although the use of different parameters makes them seem different. Essentially the plots in \cite{Donevaetal} are cross sections of our surface with surfaces of $f=\textrm{const}$ (which are not identical to the planes $j=\textrm{const}$). The change from the picture given in \cite{Donevaetal} to our picture, when passing from $f$ to $j$ (both used as measures of the rotation), is the very reason for bringing the relation between $\bar{I}$ and $\bar{Q}$ into a very precise universal form, that is EoS invariant. A comparison between our fit and previous results at zero rotation is given in Figure 7 in \cite{suppl}.

Thus, the universal relation between the moments of NSs translates to a generalized $\bar{I}-\bar{Q}$ invariance that holds for arbitrary rotation.

{\it Breaking the degeneracy.---}
%
The universal relations presented in the previous paragraphs should not be mistakenly regarded as a conclusion that the specific EoS used to describe the interior of a neutron star is irrelevant to the structure and the physical properties of the star in general; it is well established that different EoSs result in different masses and radii. The relations so far described, are referred to the reduced moments, not the moments themselves, where the principal scale, the mass $M$, has been factored out. Thus, the introduction of a scale to our description would make the degeneracy between different EoSs disappear. A choice for such a scale could be either the mass or the rotation frequency $f$ of the NS. 

\begin{figure}[t]
\vspace{-15pt}
              \includegraphics[width=.45\textwidth]{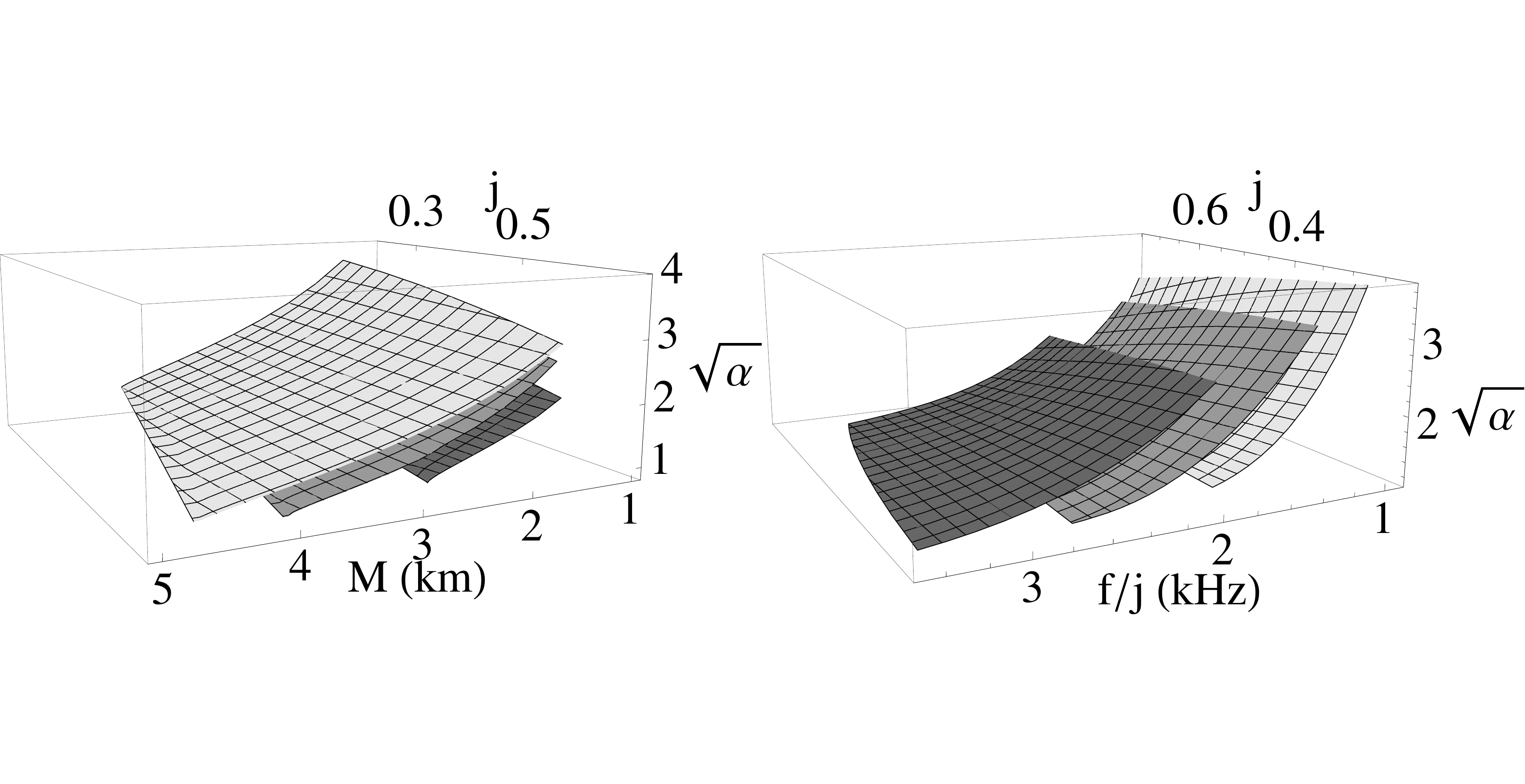}
\vspace{-20pt}
\caption{Plots of the surfaces of three different EoSs in the parameter spaces of $(M,j,\sqrt{a})$ (left) and $(j,f/j,\sqrt{a})$ (right).}
\protect\label{degeneracy}
\end{figure}

Indeed, a plot of the models in the parameter space $(M,j,\sqrt{a})$ and $(j,f/j,\sqrt{a})$ shows that different EoSs correspond to different surfaces. This can be seen in Figure \ref{degeneracy} where we have plotted for example the surfaces for the EoSs L, APR, and A (see \cite{suppl}). These surfaces can be fitted using a function of the form, 
\be
\sqrt{a}=A_1+ A_2 (\xi-\xi_0) + A_3 (\xi-\xi_0)^2, \label{fiteq}\ee
where, $A_2=B_1+B_2 j $, $A_3=C_1+C_2 j $, $j$ is the spin parameter, $\xi_0$ is a constant, and $\xi$ can be either the ratio of the frequency to the spin parameter $f/j$ in kHz or the mass $M$ in km. The results of these fits are shown in Tables III and IV in \cite{suppl}. 

Thus one could use the breaking of the degeneracy between the various EoSs by the mass or the rotation frequency in order to probe the properties of nuclear matter inside NSs. Specifically, a measurement of the first three moments of a NS (for example, see \cite{PappasQPOs,psaltis2,psaltis3,psaltis4}), could constrain the NS's EoS.

{\it Acknowledgements.---}
This work was initiated while G. Pappas was based at the section of Theoretical Astrophysics, IAAT, Eberhard Karls University of T\"ubingen. We would like to thank D. Doneva for providing some of the nuclear matter EoSs and N. Stergioulas for providing us with access to his numerical code. We also thank T. Sotiriou, I. Vega and N. Yunes for comments on the manuscript. G. Pappas acknowledges financial support from the European Research Council under the European Union's Seventh Framework Programme (FP7/2007-2013) / ERC grant agreement n. 306425 "Challenging General Relativity" and the DAAD scholarship number A/12/71258, and T. Apostolatos acknowledges DAAD scholarship. We would like to thank the anonymous referees for their comments that helped improve on the clarity of our presentation.


\clearpage

\section{Introduction}
\label{sec1}

In this supplement, we give a detailed account of all the data related to the neutron star (NS) models that we used in our analysis and all the results presented in the main text. In our analysis we have constructed rotating NS models using the RNS numerical code that was developed by \cite{Sterg}. The calculations and the graphical representation of the NS data were performed using MATHEMATICA.

\section{Equations of state and neutron star models}
\label{sec1}

The details of the realistic EoSs that we have used to build our NS models are presented in Table I. Apart from the various nuclear matter EoSs, we have also used the two EoSs proposed in \cite{SLB}, as they are inferred by Bayesian analysis from astrophysical observations of type I X-ray bursters with photospheric radius expansion and from thermal emission from quiescent low-mass X-ray binaries. The lower density part of these EoSs is matched to a crust of the form \cite{BPS}. We will generally refer to all these EoSs as ``the realistic" EoSs.

\begin{figure}[h]
\centering
\includegraphics[width=.4\textwidth]{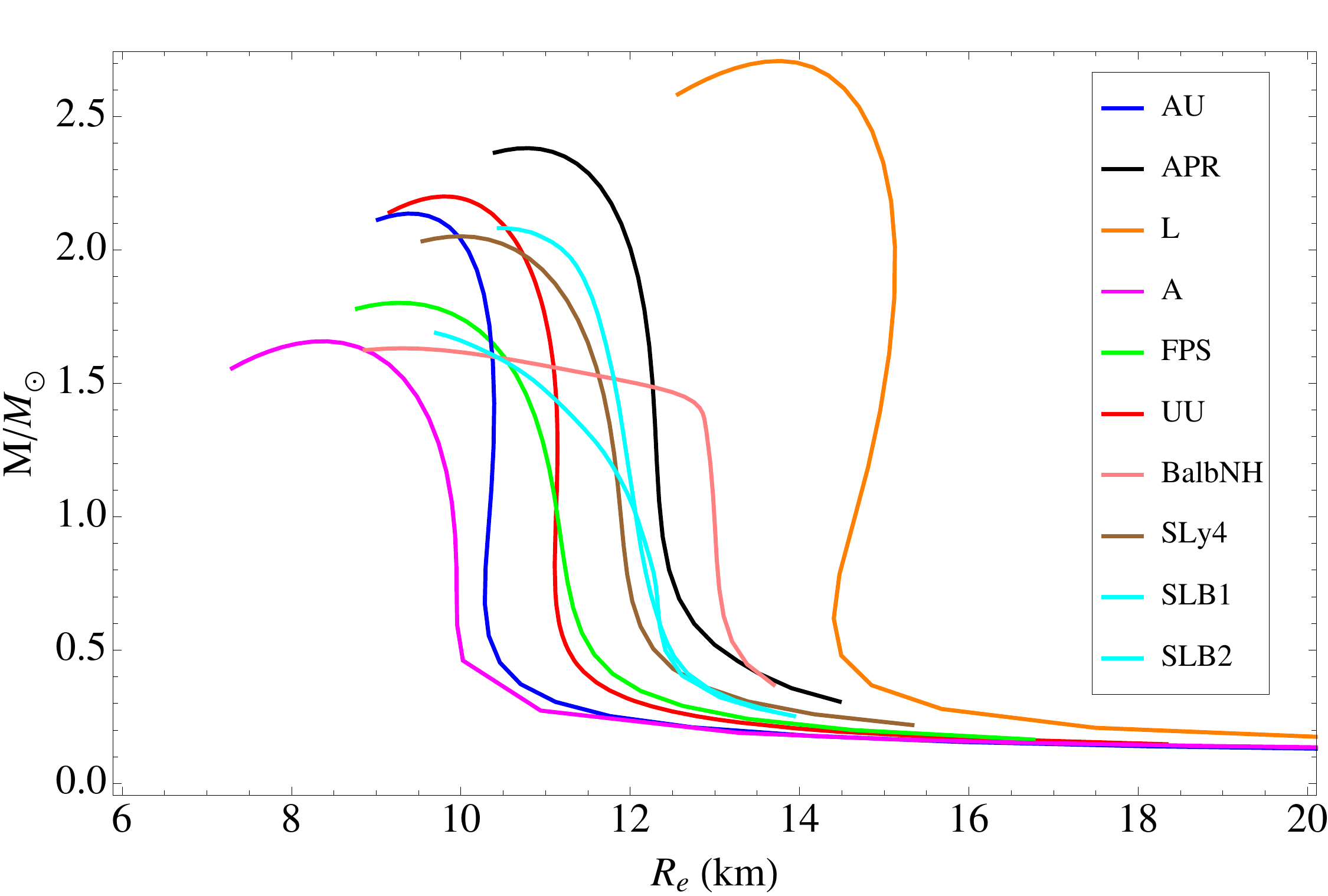}
\caption{Mass-radius relation for the various EoSs that are given in Table I.}
\protect\label{TOV}
\end{figure}

\begin{figure}[h]
\centering
\includegraphics[width=.4\textwidth]{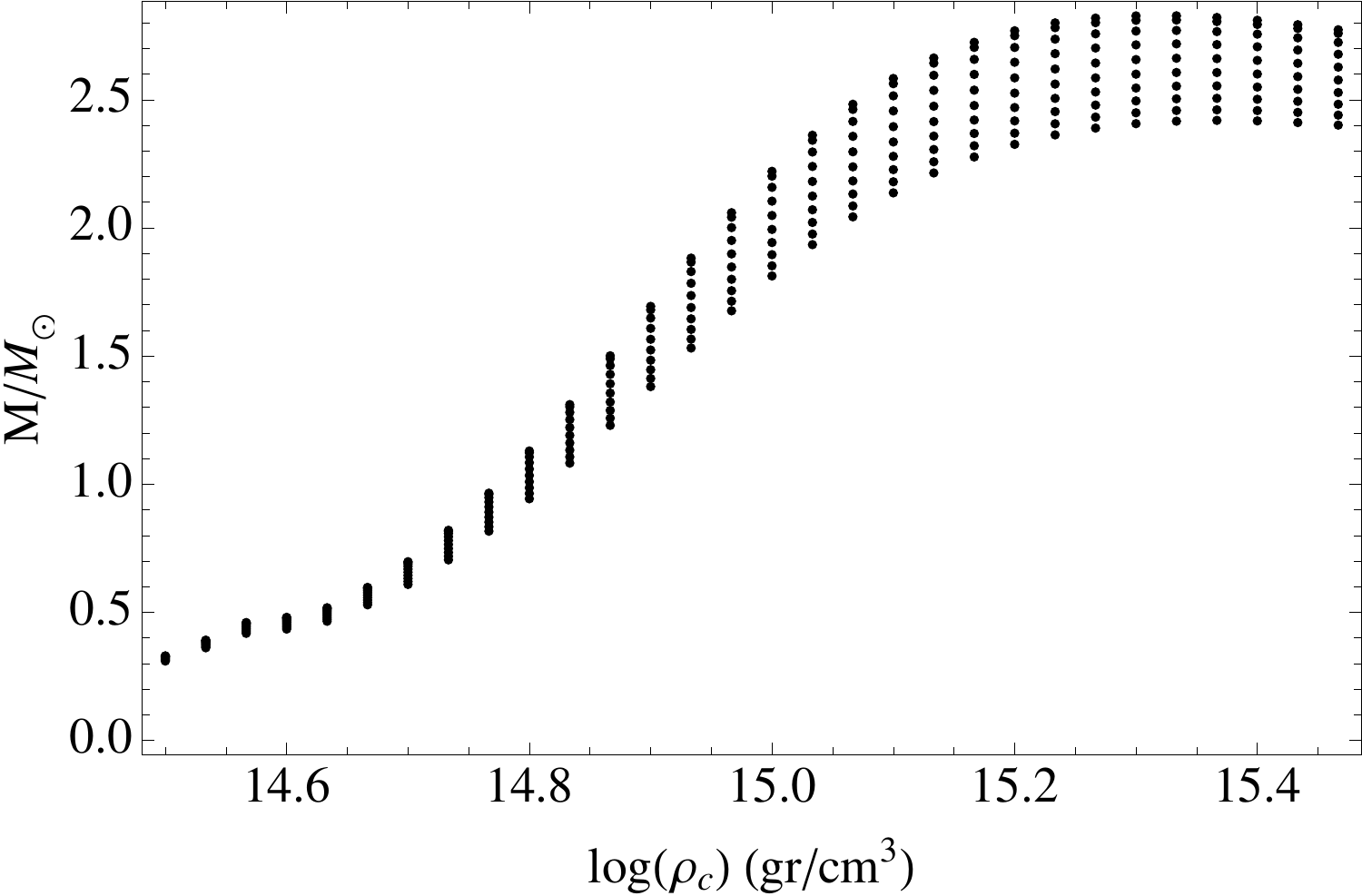}
\includegraphics[width=.4\textwidth]{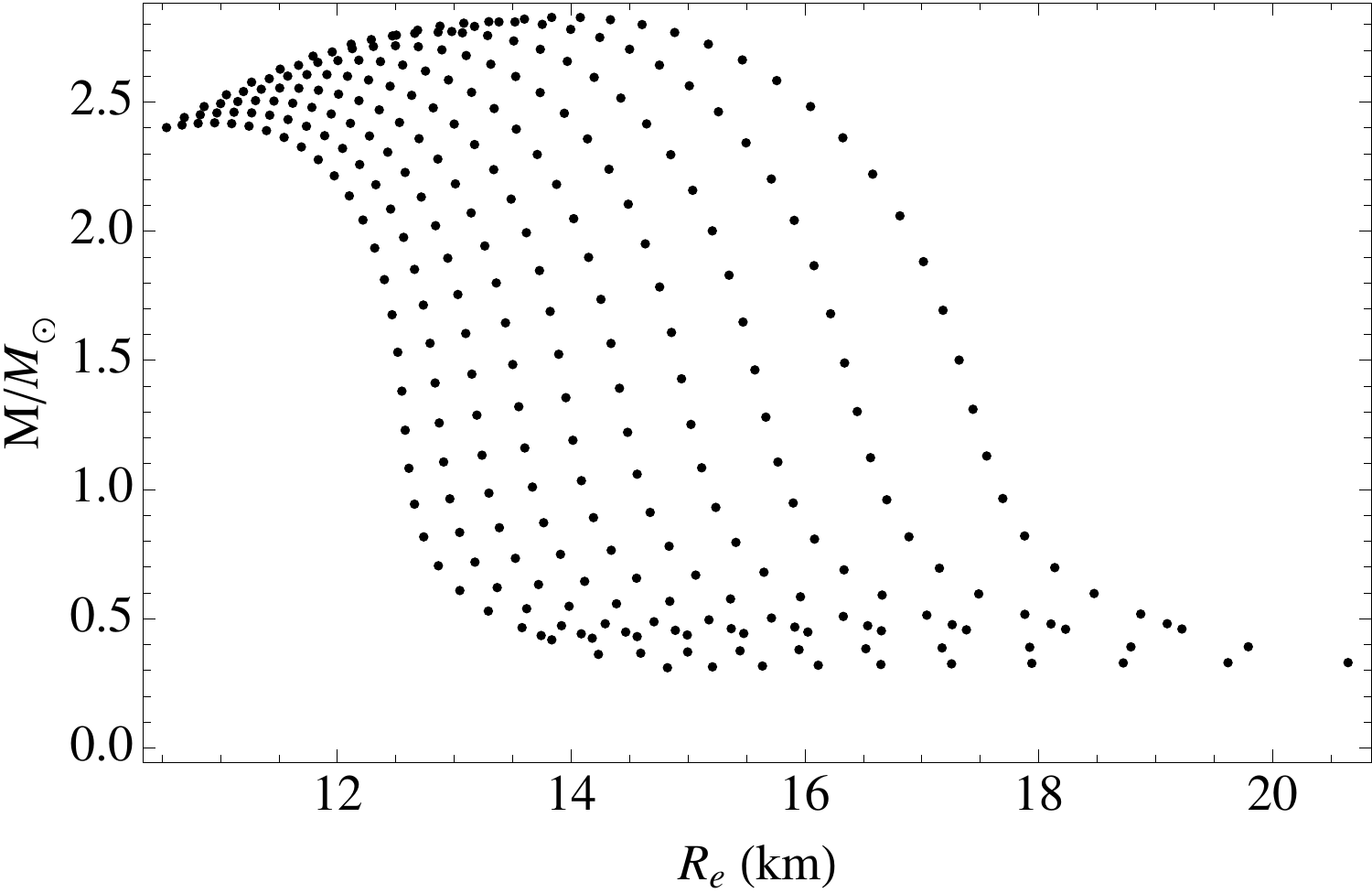}
\caption{Mass-density (top) and mass-radius (bottom) plots for the APR EoS from zero rotation up to the Kepler limit.}
\protect\label{APR}
\end{figure}

For a more schematic understanding of the features of each EoS, we give in Figure \ref{TOV} the mass-radius relations for the non-rotating NS models for each of the EoSs used in our analysis. As we can see, for all EoSs there is a region of central densities (the central density changes monotonically along the curves) for which the radius is almost independent of the central density, a behavior that is characteristic of the $n=1$ Newtonian polytrope. Each non-rotating NS sequence that we have constructed for each particular EoS is comprised by models of different central density. Then, for every such model we generated a sequence of models with increasing rotation up to the Kepler limit, i.e., the limit where the rotation is so fast that the star gets marginally disrupted. An example of these rotating sequences (for the APR EoS) is presented in Figure \ref{APR}. The upper plot shows how this sequence of rotating models, with rotation rates ranging from zero up to the Kepler limit, is generated from the non-rotating models with different central densities. The lower plot shows the mass-radius relation for all the NS models presented in the upper plot. Generally we will use models in the astrophysically relevant range of masses between the maximum stable mass, which corresponds to high densities and more compact objects, and a lower mass that is not much less than 1 solar mass, which corresponds to low density models that are located in the vertical branch of the mass-radius graph (above $0.7M_{\odot}\sim1 \textrm{km}$ depending on the EoS). 

\begin{table}[h]                                                                                                                     
\centering                    
 \protect\label{EOS}
  \caption{Details of the EoSs used.}
   \begin{ruledtabular}
 \begin{tabular}{ll}
  EoS & nuclear matter EoSs description  \\ \hline
  A	 		&	EoS A in \cite{AandB} \\ 
  AU	   		&	WFF1 (denoted as AV14+UVII in  \cite{WFF} )  \\ & matched to a crust as given in \cite{NegeleVautherin}\\
  FPS  		&	core of the type given in \cite{FPS}, \\ & crust of the type given in \cite{BPS} \\ 
  APR  		& 	EoS given in \cite{APR} with SLy4 crust \\ 
  UU	   		&	WFF2 (denoted as UV14+UVII in \cite{WFF}) \\ & matched to a crust of type  \cite{NegeleVautherin} \\ 
  L       		&     	EoS L in \cite{AandB} \\ 
  SLy4 		&  	EoS described in \cite{SLy4} \\ 
  BalbNH		&	Core as given in \cite{balb1}, Crust:  SLy4 crust \\ \hline
  		 	& 	EoSs constructed from observational constrains \\ \hline
  SLB1 		& 	$r_{ph}=R$, EoS given in \cite{SLB} matched to \\ & a crust as in \cite{BPS}  \\ 
  SLB2	 	& 	$r_{ph}\gg R$, EoS given in \cite{SLB} matched to \\ & a crust as in \cite{BPS}  \\ 
 \end{tabular}
\end{ruledtabular}
\noindent 
SLy4 = A unified crust-core equation of state, calculated using an effective S(kyrme-)Ly(on) nuclear interaction for both crust (in nucleus bulk energy) and core (in nucleon-nucleon energy) by \cite{SLy4}.
\end{table}

%

\section{Universal surfaces and fit parameters}
\label{sec2}

Here we present some more details regarding the surfaces that NSs form in the parameter space of the reduced moments and their ratios. The two types of parameter spaces that we have used are the $(x,y,z)$ space, where $x\equiv j$, $y\equiv (-q)^{1/2}$, and $z\equiv (-s_3)^{1/3}$ , and the $(j,\sqrt{a},\sqrt[3]{\beta})$ space.

\begin{figure}
\includegraphics[width=.44\textwidth]{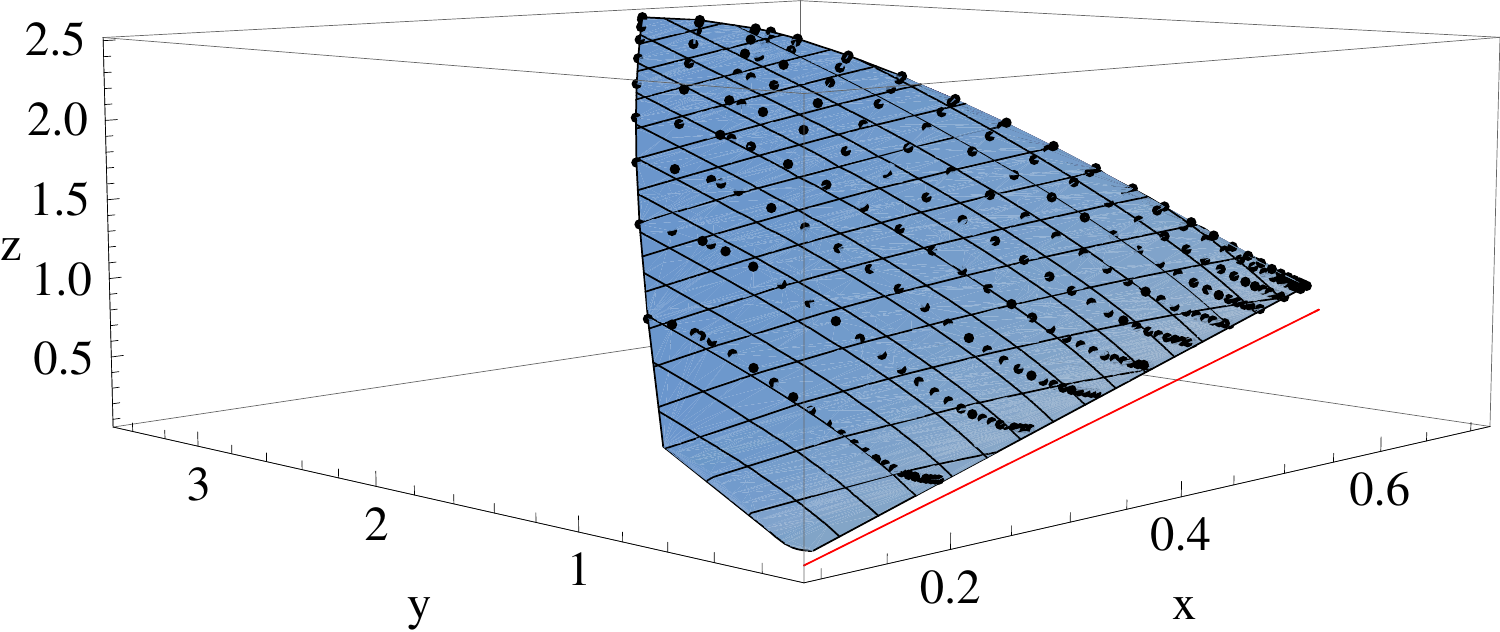} 
\includegraphics[width=.44\textwidth]{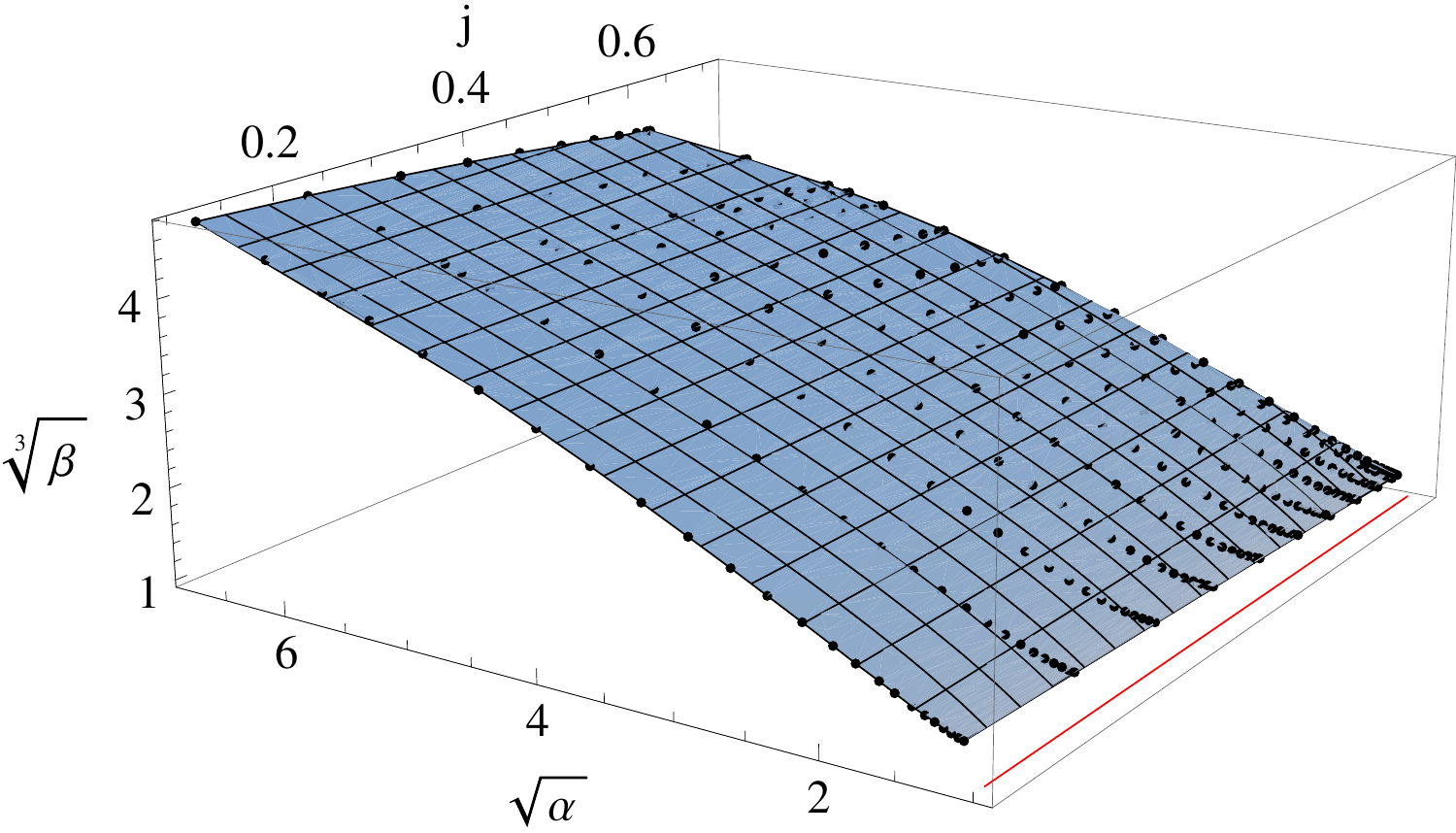}
\caption{Parameter space plots for the APR EoS. The upper plot shows the surface that is formed by the NS models in the parameter space $(x,y,z)$, along with the points of the individual NS models lying on that surface. The density increases in the azimuthal direction from left to right, while rotation increases radially. The bottom plot shows the same NS models in the parameter space $(j,\sqrt{a},\sqrt[3]{\beta})$. Now the rotation increases with increasing $j$ and the density increases with decreasing $\sqrt{a}$. The red line in both plots indicates the curve of Kerr black holes with different spin parameters $j$ (the spin parameter increases as we move away from the origin) and it is located near the high density edge of the surface.}
\protect\label{APRsurface}
\end{figure}

As an example we show the surfaces for the APR EoS in Figure \ref{APRsurface}. The upper plot shows the surface that is formed by the NS models in the parameter space of $(x,y,z)$; the individual NS models are depicted as points on this surface. The red line indicates the Kerr black hole solutions, for which we have that $-q^{Kerr}=j^2$ and $-s_3^{Kerr}=j^3$. Thus this line is parallel to the vector $\vec{r}=(1,1,1)$. We can see that the surface approaches the Kerr line along one of its edges. The models of the same central density fall along radial curves with the density increasing in the azimuthal direction from the left part of the upper plot towards the right part near the red radial line of Kerr black holes. The rotation increases along the radial direction with the $(0,0,0)$ point corresponding to the non-rotating models, while the furthest points from the origin correspond to the maximally rotating models. The mass increases as the central density increases, as long as we are on the branch of stable NS models. On the other hand, the mass increases with increasing rotation, for the same central density (see Figure \ref{APR}). The bottom plot shows the corresponding surface in $(j,\sqrt{a},\sqrt[3]{\beta})$.

\begin{figure}
\centering
\includegraphics[width=.5\textwidth]{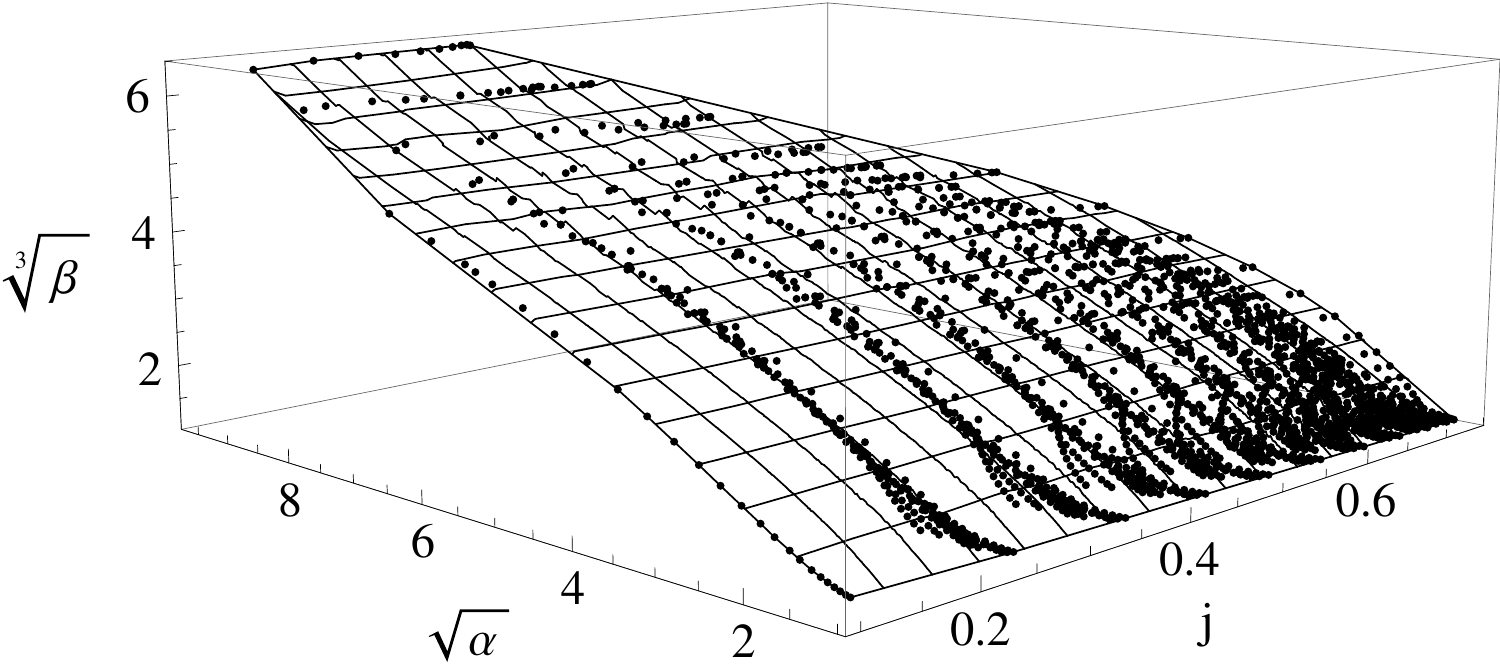}
\caption{Parameter space plot for all the realistic EoSs. All the models from all the EoSs fall on one very well defined surface.}
\protect\label{UniSurfaceSupl}
\end{figure}

From the bottom plot of Figure \ref{APRsurface}, we can see that $\sqrt[3]{\beta}$ is almost independent of $j$. As a fitting function we have chosen a function of the form, 
\be \sqrt[3]{\beta}=B (\sqrt{a})^{\nu}. \nn \ee
The best fitting parameters we have found are, $B=1.16$, and $\nu=0.75$, which are close to the parameters describing the universal fit for all the EoSs that is given in the main text and in Table II here. We can obtain a better fit if we try a function of the form $ \sqrt[3]{\beta}=A+B (\sqrt{a})^{\nu}$, where the coefficients are in this case, $A=-0.97$, $B=2.02$, and $\nu=0.54$. The behavior of the other EoSs of Table I is similar and the surface formed by all the EoSs is shown in Figure \ref{UniSurfaceSupl}. 

\begin{figure}
\centering
\includegraphics[width=.45\textwidth]{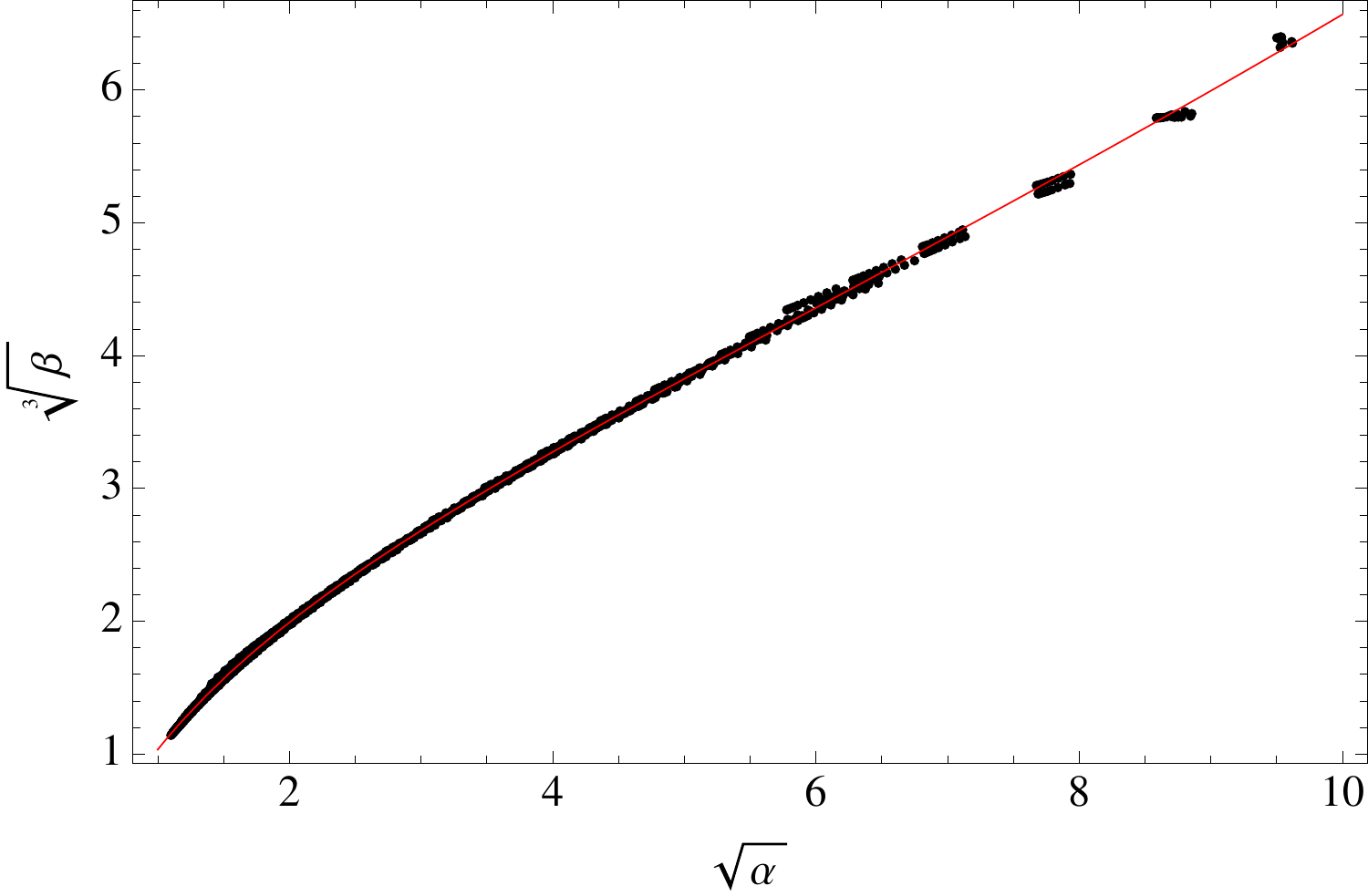}
\caption{Projection of all the data points for the NS models and their fitting surface onto the $(\sqrt{a},\sqrt[3]{\beta})$ plane. The fitting function shown is a double power law of the form, $ \sqrt[3]{\beta}=A+B_1 (\sqrt{a})^{\nu_1}+B_2 (\sqrt{a})^{\nu_2}$, that gives the most accurate fit.}
\protect\label{UniSurfaceSupl2}
\end{figure}

Additionally, in Figure \ref{UniSurfaceSupl2} we show the projection of the surface of realistic EoSs on the $(\sqrt{a},\sqrt[3]{\beta})$ plain. All models, independent of the EoS, fall almost on a line, which justifies our choice to describe the surface with a function independent of $j$.  One could get an even better fit than those given in Table II for the surface of realistic EoSs, if a more complicated fitting function had been used. If for example we use a double power law of the form, $ \sqrt[3]{\beta}=A+B_1 (\sqrt{a})^{\nu_1}+B_2 (\sqrt{a})^{\nu_2}$, we can fit the surface with an accuracy better than 2\% for all $\sqrt{a}$  with the parameters, $A=-4.82$, $B_1=5.83$, $\nu_1=0.205$, $B_2=0.024$, and $\nu_2=1.93$. This accuracy of better than 2\% gives as an estimate of how well all the models fall on one surface. This is the red curve shown in Figure \ref{UniSurfaceSupl2}.

For the comparison between the realistic EoSs and the polytropic EoSs we have constructed sequences of polytropic models with polytropic indices, $n=0.5,1,1.5,2$, and $2.5$. The fitting parameters for the surfaces in the space of $(j,\sqrt{a},\sqrt[3]{\beta})$ for these EoSs are given in Table II together with the fitting parameters for the surface of all the realistic EoSs of Table I and the specific example of APR EoS.

\begin{table}
 \caption{Fitting parameters for the surfaces of the models constructed with polytropic EoSs and the surface of the models constructed using the EoSs given in Table I. The coefficients correspond to a function of the form, $\sqrt[3]{\beta}=A+B (\sqrt{a})^{\nu}$, while the ``$-$" indicates a fitting choice where the specific parameter is not used. For the polytropic EoSs the fit is better than 3\% if we include the parameter $A$ and better than 5\% if we don't include it. For the surface of the realistic EoSs  the fit is better than 5\% with $A$, while without $A$ it ranges within 5-10\% for $\sqrt{a}<1.2$. This is the region of the maximum mass NSs, where $\sqrt{a}$ is close to 1.}
 \label{fit}
 \centering
   \begin{ruledtabular}
 \begin{tabular}{@{}lccc}
  EoS & $A$ & $B$ & $\nu$  \\ \hline
  APR EoS & $-$ & 1.16 & 0.75 \\
      		 & -0.97 & 2.02 & 0.54 \\ \hline
 nuclear matter + observational & $-$ & 1.17 & 0.74 \\
						& -0.36 & 1.48 & 0.65 \\ \hline
 $n=0.5$ polytrope & $-$ & 1.27 & 0.68 \\
 			      & -0.45 & 1.56 & 0.63 \\ \hline
 $n=1$ polytrope & $-$ & 1.27 & 0.7 \\
                          & -0.71 & 1.83 & 0.58 \\ \hline
 $n=1.5$ polytrope & $-$ & 1.36 & 0.68 \\
  				& -0.46 & 1.68 & 0.62 \\ \hline
 $n=2$ polytrope & $-$ & 1.44 & 0.68 \\
 				 & -0.4 & 1.71 & 0.63 \\ \hline
 $n=2.5$ polytrope & $-$ & 1.56 & 0.67 \\
 			 & 0.03 & 1.55 & 0.67 \\
 \end{tabular}
   \end{ruledtabular}
\end{table}

\section{I-Love-Q}
\label{sec2_5}

\begin{figure}[h]
\centering
\includegraphics[width=.5\textwidth]{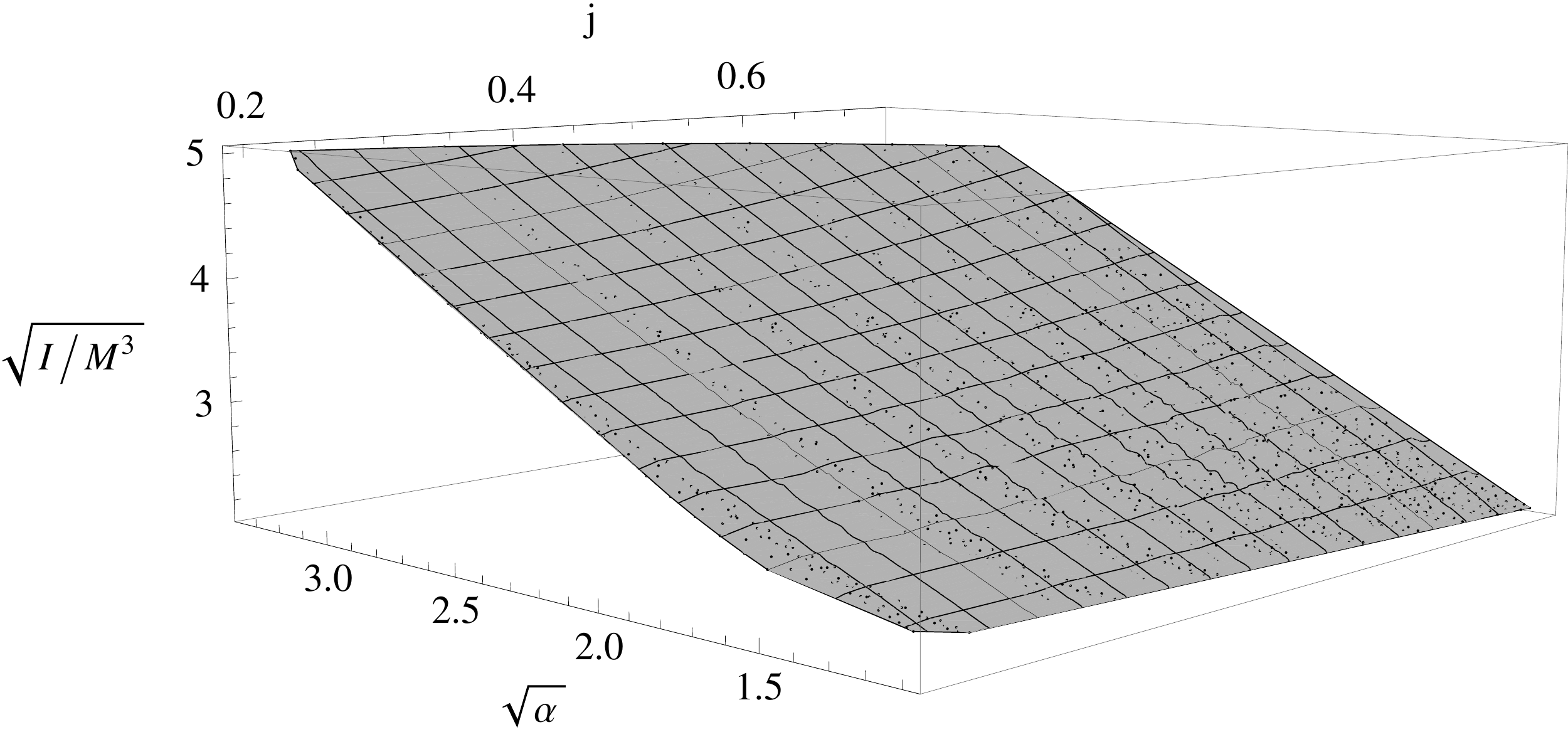}
\includegraphics[width=.5\textwidth]{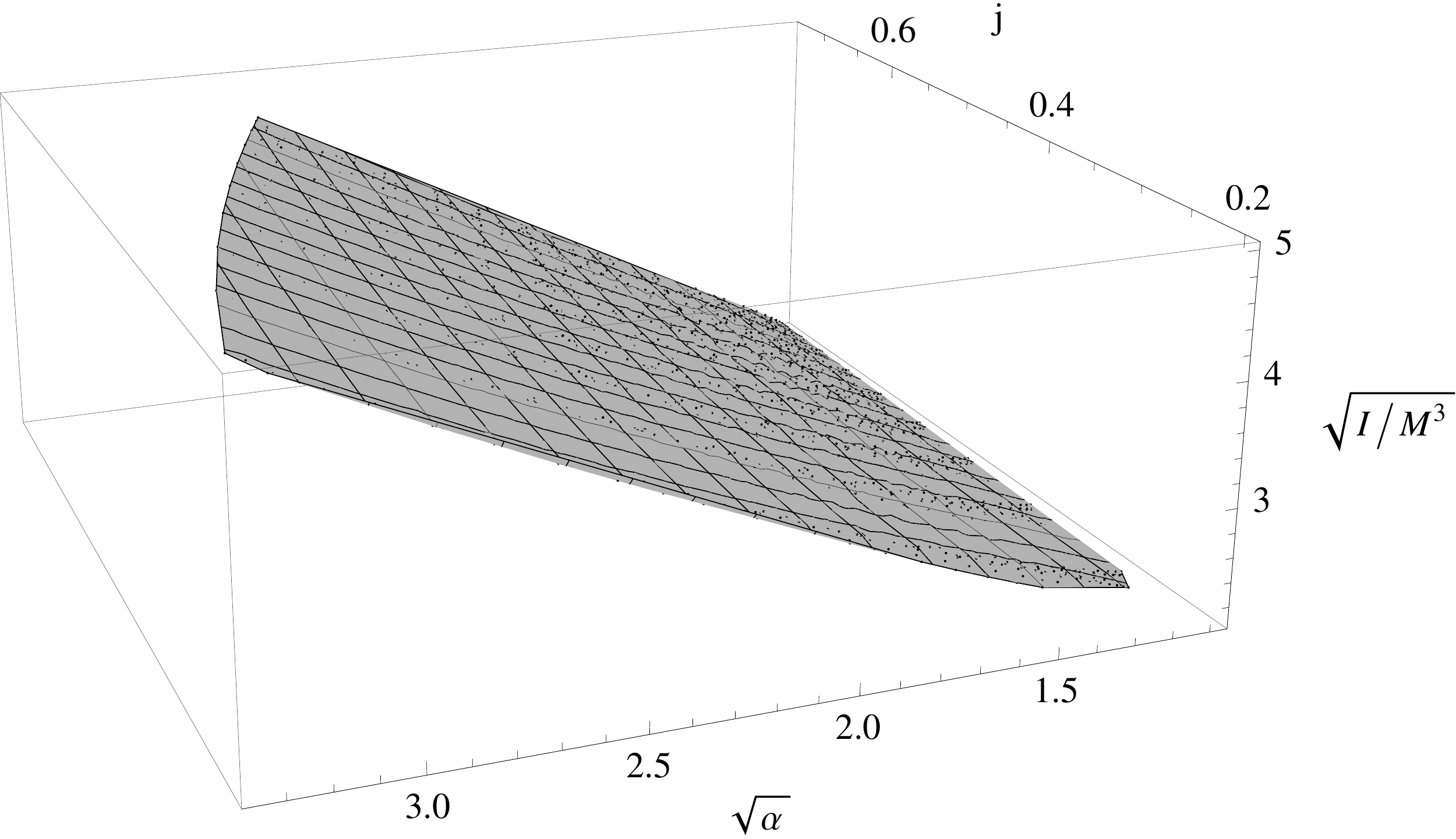}
\includegraphics[width=.5\textwidth]{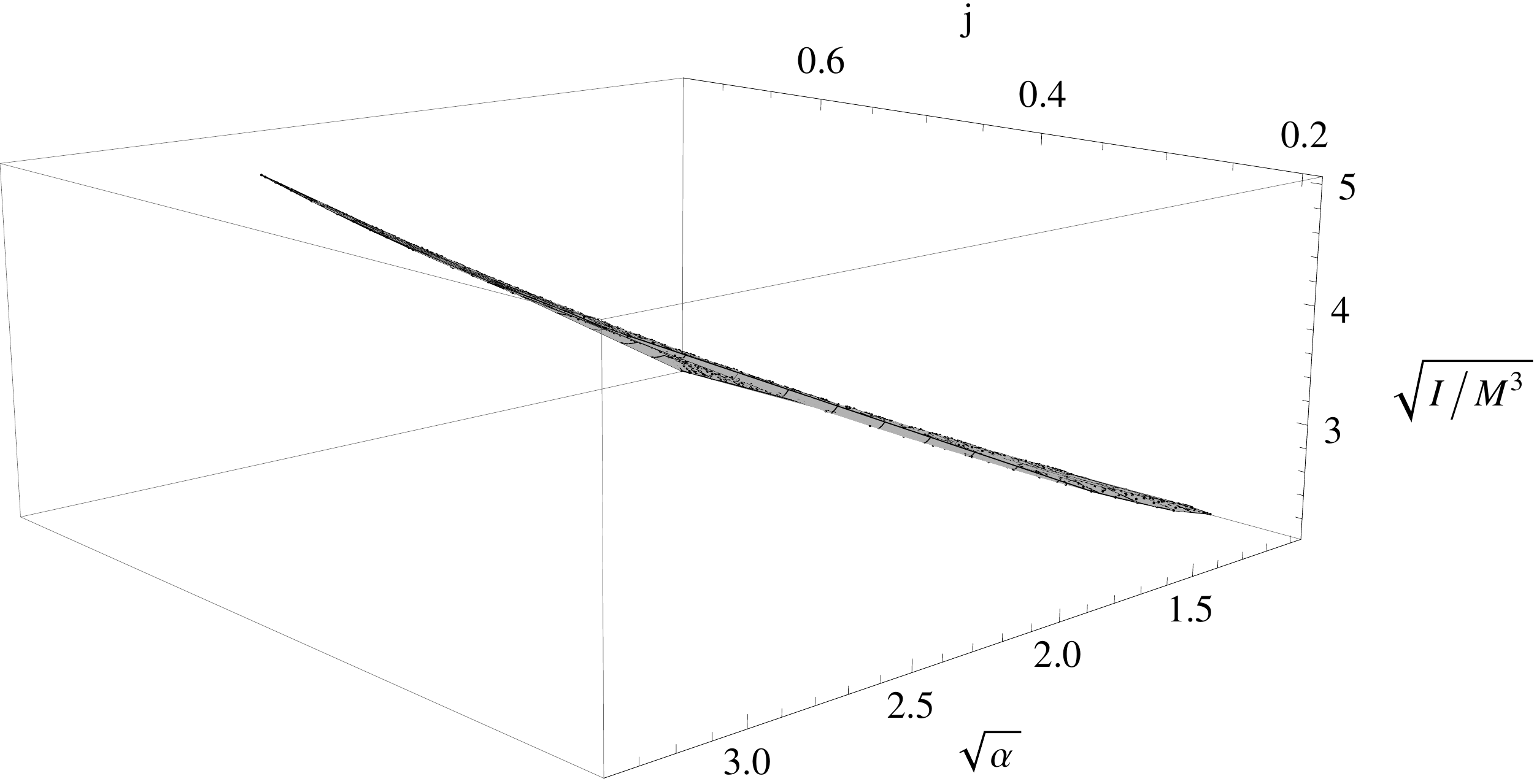}
\caption{The plot shows the surface that the NSs form (for all the realistic EoSs) in the parameter space $(j,\sqrt{a},\sqrt{I/M^3})$. The surface is shown from three different view points in order to demonstrate that all the models lie on one surface.  
}
\protect\label{CompSurface2}
\end{figure}

\begin{figure}
\centering
\includegraphics[width=.45\textwidth]{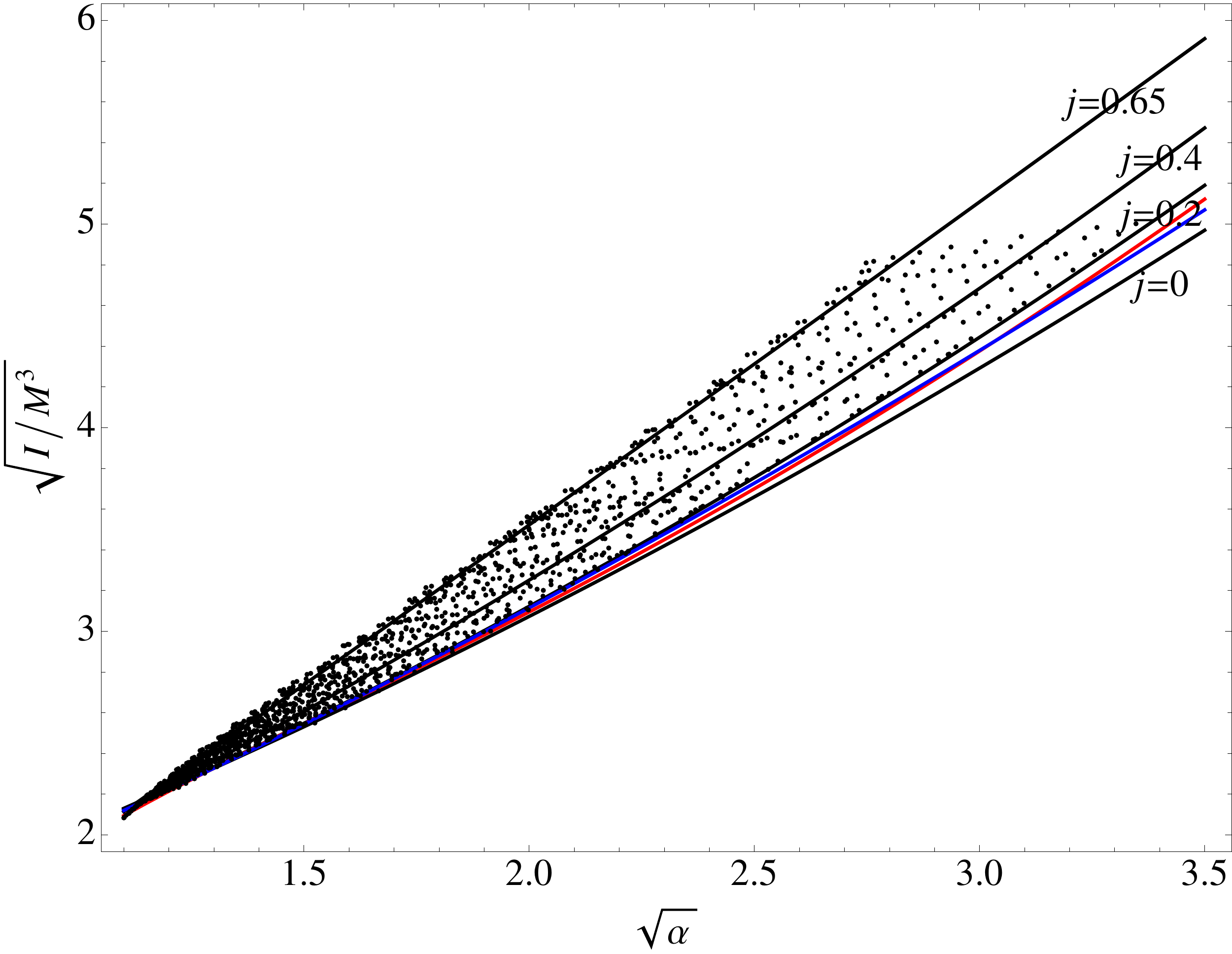}
\caption{The plot shows the projection of all NS models in the parameter space $(j,\sqrt{a},\sqrt{I/M^3})$ onto the $j=0$ plane. The plot shows also the curves obtained by (\ref{iqfit}) when the spin parameter is fixed at some specific values. The plot also shows the fit presented in \cite{Yagi-Yunes} in red and the fit given in \cite{Donevaetal} for zero rotation in blue. Our own zero rotation curve deviates from the other two by at most 2\%.
}
\protect\label{CompSurface3}
\end{figure}

Here we present the results regarding the depiction of all NS models in the parameter space of $(j,\sqrt{a},\sqrt{I/M^3})$ for comparison with the results of \cite{Yagi-Yunes,Donevaetal}. Figure \ref{CompSurface2} shows the surface that is formed by NSs constructed with all the realistic EoSs. Again in this case all the EoSs fall on one surface. Figure \ref{CompSurface3} shows a projection of the data points, that form the previous surface, on the plane of $(\sqrt{a},\sqrt{I/M^3})$. The latter Figure makes apparent the connection between our results and those presented in \cite{Donevaetal}. The lines plotted are cross-sections, of planes of specific values of the spin parameter $j$, with the fitting function for the surface given as,
\be
\sqrt{I/M^3}=A_1+ A_2 (\sqrt{a}-\xi_0) + A_3 (\sqrt{a}-\xi_0)^2, \label{iqfit}\ee
where, $A_2=B_1+B_2 j +B_3 j^2$, $A_3=C_1+C_2 j +C_3 j^2$, $j$ is the spin parameter, and $A_1=2.16$, $\xi_0=1.13$, $B_1=0.97$, $B_2=-0.14$, $B_3=1.60$, $C_1=0.09$, $C_2=0.23$, and $C_3=-0.54$. These parameters give a fit that is better than 1\%. Again this can be considered as a measure of how well the models fall on a surface.

\section{Breaking the degeneracy and fit parameters}
\label{sec3}

To break the degeneracy between the different EoSs we need to introduce a scale to our description. We do that with two choices for the scale, the mass of the star and the rotation frequency $f$ of the star. 

\begin{figure}
              \includegraphics[width=.5\textwidth]{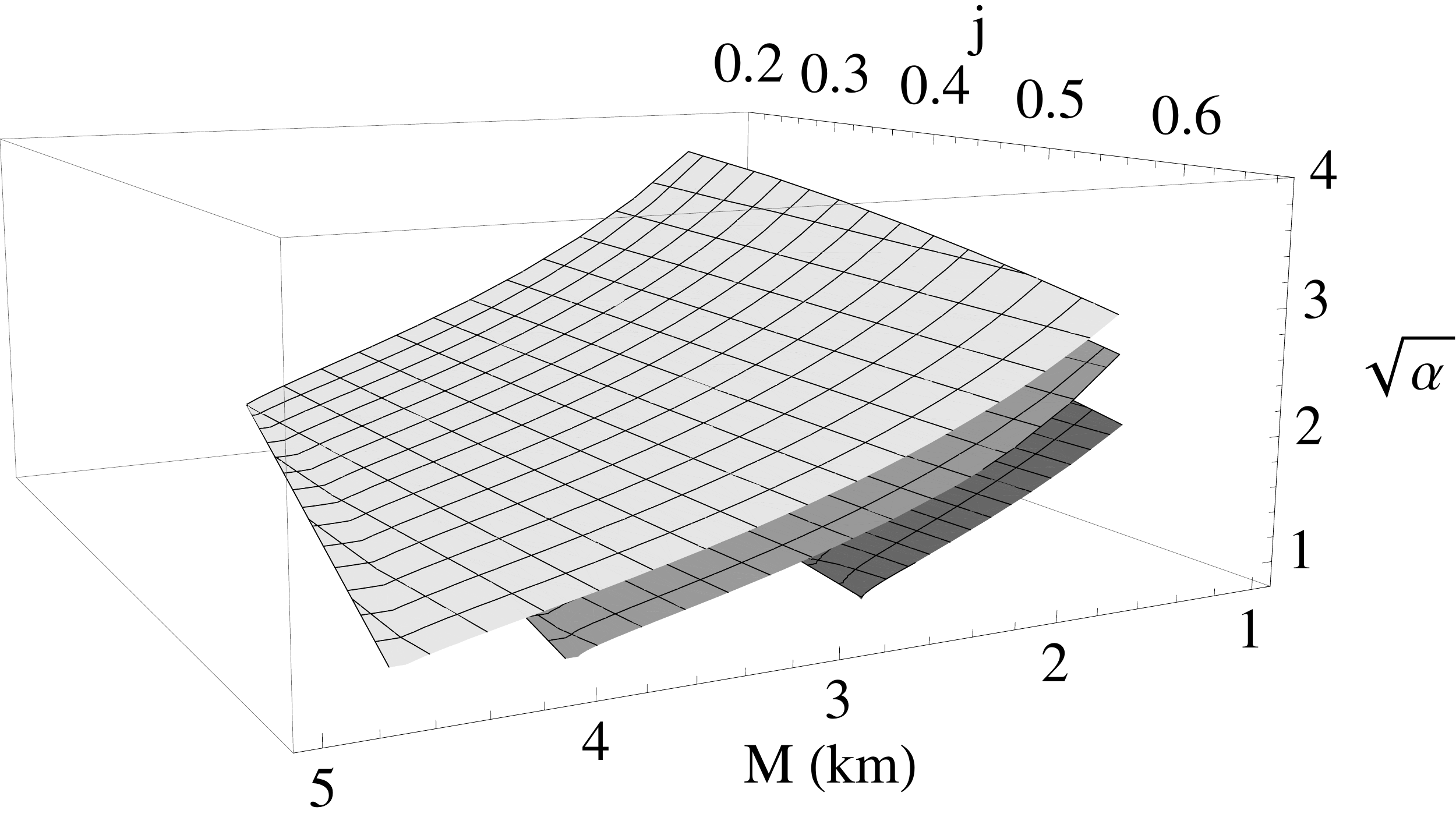}
                \includegraphics[width=.5\textwidth]{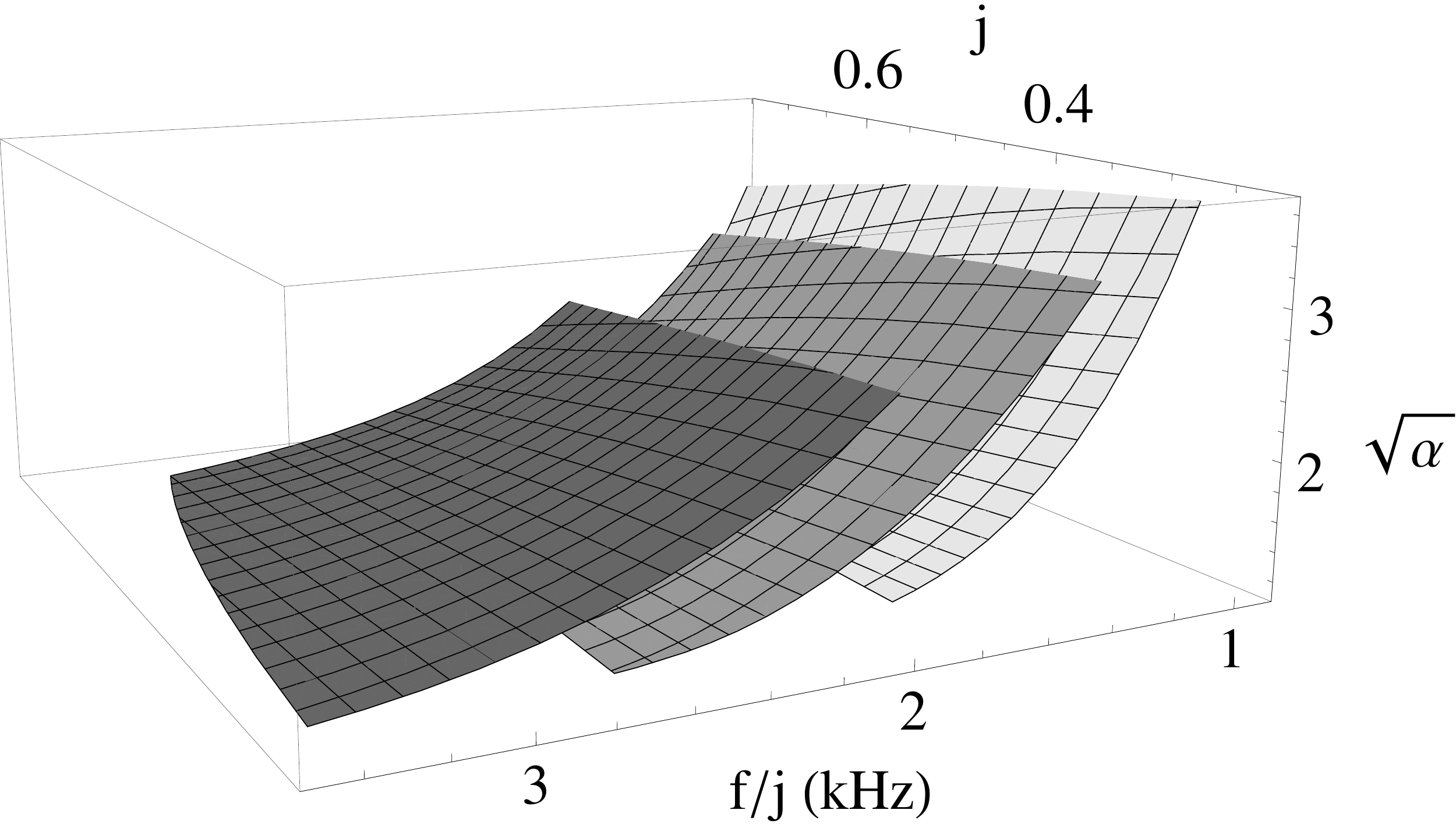}
\caption{Plots of the surfaces of three different EoSs in the parameter spaces $(M,j,\sqrt{a})$ (top) and $(j,f/j,\sqrt{a})$ (bottom). In the top plot from left to right the surfaces of EoSs L, APR, and A are depicted, while in the bottom plot the corresponding surfaces for the same EoSs are shown from right to left. These are the same plots as the ones shown in Figure 3 of the main text.
}
\protect\label{degeneracy2}
\end{figure}

Plots of the models in the parameter space of  $(M,j,\sqrt{a})$ and $(j,f/j,\sqrt{a})$ are depicted in Figure \ref{degeneracy2}, where we have indicatively plotted the surfaces for the EoSs L, APR and A. All surfaces can be fitted with a function of the form, 
\be
\sqrt{a}=A_1+ A_2 (\xi-\xi_0) + A_3 (\xi-\xi_0)^2, \nn\ee
where, 
\be A_2=B_1+B_2 j, \; A_3=C_1+C_2 j, \nn \ee 
$j$ is the spin parameter, $\xi_0$ is a constant, and $\xi$ can be either the ratio of the frequency to the spin parameter, $f/j$ (in kHz) in the one case, or the mass $M$ (in km) in the other. The results of these fits are shown in Table III for the frequency and in Table IV for the mass. 

The asterisk in Table IV next to the EoS BalbNH is indicating a problematic fit of this EoS with the fitting function that we chose, due to its different behavior in the mass-radius diagram compared to the rest of the EoSs; so, no fitting parameters are given for this EoS. More specifically, the problem with this particular EoS was that the parameters of the fit would give a function that would have unreasonable behavior for lower masses and zero rotation.

%
\begin{table}[h]
 \caption{Fitting parameters for the surfaces of the EoSs shown in Table I in the $(j,f/j,\sqrt{a})$ space. For all fits, the accuracy in calculating $\sqrt{a}$ is better than 2-5\%. The last two columns show the maximum values for the frequency over $j$ (in kHz) and the spin parameter for every EoS so as to indicate within which range the fitting function can be used.}
 \label{degeneracyfit1}
 \centering

\begin{tabular}{lcccccccc}\hline
 EoS & $A_1$ & $\xi_0$ & $B_1$ & $B_2$ & $C_1$ & $C_2$ & $(f/j)_{max}$ & $j_{max}$ \\ 
  & & & & & & & (kHz) & \\  \hline
 APR & 1.2 & 2.69 & -0.91 & 2.07 & 0.95 & 0.19  & 2.71 & 0.71 \\
 AU & 1.17 & 3.07 & -0.87 & 1.85 & 0.82 & 0.06 & 3.1 & 0.73 \\
 A & 1.31 & 3.59 & -0.34 & 1.17 & 0.54 & -0.02 & 3.49 & 0.68 \\
 UU & 1.2 & 2.94 & -0.81 & 1.88 & 0.82 & 0.14 & 2.98 & 0.72 \\
 FPS & 1.35 & 3.15 & -0.51 & 1.61 & 0.66 & 0.15  & 3.14 & 0.67 \\
 L & 1.39 & 1.99 & -1.39 & 3.65 & 1.84 & 0.98 & 1.98 & 0.74 \\
 BalbNH & 1.58 & 2.86 & -0.5 & 1.91 & 0.54 & 0.44 & 3.02 & 0.69 \\
 SLy4 & 1.28 & 2.94 & -0.56 & 1.6 & 0.78 & 0.06 & 2.91 & 0.68 \\
 SLB1 & 1.45 & 3.15 & -0.47 & 1.56 & 0.5 & 0.23 & 3.03 & 0.68 \\
 SLB2 & 1.29 & 2.78 & -0.86 & 2.06 & 0.79 & 0.28 & 2.69 & 0.68 \\
 \end{tabular}
%
\end{table}

\begin{table}
 \caption{Fitting parameters for the surfaces of the EoSs shown in Table I in the $(M,j,\sqrt{a})$ space. For all the fits, the accuracy in calculating $\sqrt{a}$ is better than 3-5\%. The ``$-$" indicates that this parameter is not used. The last two columns show the maximum values for the mass (in km) and the spin parameter for every EoS so as to indicate within which range the fitting function can be used.}
 \label{degeneracyfit2}
 \centering
\begin{tabular}{lcccccccc}\hline
 EoS & $A_1$ & $\xi_0$ & $B_1$ & $B_2$ & $C_1$ & $C_2$  & $M_{max}$ & $j_{max}$ \\ 
   & & & & & & &(km) &  \\ \hline
 APR & 5.12 & $-$ & -1.6 & -0.21 & 0.13 & 0.08 & 4.17 & 0.71 \\
 AU & 4.87 & $-$ & -1.7 & -0.23 & 0.16 & 0.1 & 3.78 & 0.73 \\
 A & 4.38 & $-$ & -1.01 & -0.7 & -0.12 & 0.37 & 2.89 & 0.68 \\
 UU & 5.07 & $-$ & -1.7 & -0.25 & 0.15 & 0.1 & 3.86 & 0.72 \\
 FPS & 5.14 & $-$ & -1.57 & -0.54 & 0.05 & 0.27 & 3.13 & 0.67 \\
 L & 5.35 & $-$ & -1.33 & -0.31 & 0.08 & 0.09 & 4.83 & 0.74 \\
 BalbNH* & $-$ & $-$ & $-$ & $-$ & $-$ & $-$ & 2.75 & 0.69 \\
 SLy4 & 4.94 & $-$ & -1.4 & -0.39 & 0.06 & 0.18 & 3.57 & 0.68 \\
 SLB1 & 5.48 & $-$ & -1.51 & -0.73 & -0.06 & 0.4 & 2.9 & 0.68 \\
 SLB2 & 5.28 & $-$ & -1.72 & -0.29 & 0.14 & 0.13 & 3.65 & 0.68 \\
 \end{tabular}
%
\end{table}
%

\clearpage

\end{document}